\documentclass[prl,twocolumn,balancelastpage]{revtex4}
\usepackage{amsmath}
\usepackage{graphicx}

\input{tcilatex}

\begin{document}

\title{Ultrasonic cavity solitons}
\author{I. P\'{e}rez--Arjona$^{1}$}
\author{V. J. S\'{a}nchez--Morcillo$^{1}$}
\author{G. J. de Valc\'{a}rcel$^{2}$}
\affiliation{$^{1}$Departament de F\'{\i}sica Aplicada, Universitat Polit\`{e}cnica de Val%
\`{e}ncia, Crta. Natzaret-Oliva s/n, 46730 Grau de Gandia, Spain}
\affiliation{$^{2}$Departament d'\`{O}ptica, Universitat de Val\`{e}ncia, Dr. Moliner 50,
46100-Burjassot, Spain}

\begin{abstract}
We report on a new type of localized structure, an ultrasonic cavity
soliton, supported by large aspect-ratio acoustic resonators containing
viscous media. The spatio-temporal dynamics of this system is analyzed on
the basis of a generalized Swift-Hohenberg equation, derived from the
microscopic equations under conditions close to nascent bistability. These
states of the acoustic and thermal fields are robust structures, existing
whenever a spatially uniform solution and a periodic pattern coexist. An
analytical solution for the ultrasonic cavity soliton is also presented.

\end{abstract}

\maketitle

Many systems in nature, when driven far from equilibrium, can self-organize
giving rise to a large variety of patterns or structures. Although studied
intensively for most of the last century, it has only been during the past
thirty years that pattern formation has emerged as an own branch of science 
\cite{Cross93}. One of the most relevant features of pattern formation is
its universality: systems with different microscopic descriptions frequently
exhibit similar patterns on a macroscopic level. This universal character of
pattern formation is evidenced when the microscopic models are reduced,
under given assumptions, to simpler equations describing the evolution of a
single variable, the so called order parameter \cite{Newell93}. Order
parameter equations are usually obtained near some critical point, and
belong to few and well known classes, such as the Ginzburg-Landau or the
Swift-Hohenberg equations and their variants. They are often based on system
symmetries and are independent of the microscopic differences among systems,
providing a theoretical framework to understand the origins of
non-equilibrium pattern formation \cite{Cross93}.

This approach to pattern formation has been extensively applied to nonlinear
optical cavities \cite{NM92,Mandel97,Springer03}, such as lasers, optical
parametric oscillators or Kerr (cubically nonlinear) resonators, where light
in the transverse plane of the cavity has been shown to develop patterns
with different symmetries (rolls, hexagons, and also quasi-patterns) as well
as cavity solitons (CS). The latter correspond to localized solutions often
resulting from a bistability between two stationary, spatially extended
states of the system, in the form of either self-trapped switching waves
between two homogeneous states, or isolated pattern elements embedded in a
homogeneous background. Optical CSs are particularly interesting objects
because of their potential use as memory bits in optical information
processing systems \cite{Barland02}. However, despite the existing analogies
between optics and acoustics, acoustical resonators in the nonlinear regime
have been much less explored than their optical counterparts. Furthermore,
pattern formation studies in acoustics are almost lacking. The main reason
lies in the weak dispersion of sound in common homogeneous media, which is
responsible for the growth of higher harmonics, leading to wave distortion
and shock formation. These effects are absent in optics, which is dispersive
in nature. However, in some special cases it is still possible to avoid the
nonlinear distortion and recover the analogies \cite{Bunkin86}. It is, for
example, the case of sound beams propagating in viscous media characterized
by a strong absorption (e.g. glycerine), where sound velocity depends of
fluid temperature, resulting in an additional nonlinearity mechanism of
thermal origin. For the majority of fluids, temperature variations induced
by an intense acoustic field result in a decrease of sound velocity, leading
to a self-focusing of the beam. In the case of viscous fluids the
characteristic length of self-focusing effects is much shorter than the
corresponding to the developement of shock waves \cite{Bunkin94}. Also, high
frequency components are strongly absorbed in such media, so in practice the
use of a quasimonochromatic (optical) description for wave propagation is
justified.

In a viscous medium sound propagates with a speed $c$ that depends
signifficantly on temperature, $c=c_{0}\left( 1-\sigma T^{\prime }\right) $,
where $c_{0}$ is the speed of sound at some equilibrium (ambient)
temperature, $T^{\prime }$ denotes the variation of the medium temperature
from that equilibrium due to the intense acoustic wave, and $\sigma $ is the
parameter of thermal nonlinearity. The propagation of sound in such a medium
has been shown \cite{Naugolnykh, Bunkin94} to be described in terms of two
coupled equations for pressure, $p^{\prime }$, and temperature, $T^{\prime }$%
, deviations. These equations have been used to address problems such as
self-focusing and self-transparency of sound \cite{Bunkin94}. They have been
also the basis for the analysis of temporal dynamic phenomena in acoustic
resonators \cite{Lyakhov93}. In this case, a viscous fluid is bounded by two
flat and parallel reflecting surfaces. One of the surfaces, vibrating at a
frequency $f$, is an ultrasonic source providing the external forcing.
Previous studies on this system \cite{Lyakhov93,Andreev86} have reported, in
the frame of the plane-wave approximation, bistability and complex temporal
dynamics, in good agreement with the corresponding experiments. In this
Letter we extend the previous model considering the effect of sound
diffraction and temperature diffusion in a large aperture resonator. These
effects, which are responsible of the spatial coupling, can play an
important role when the Fresnel number of the resonator $\mathcal{F}%
=l^{2}/\lambda L>>1$ (being $l$ and $L$ its transverse and longitudinal
dimensions, respectively).

The intracavity pressure field is decomposed into two counterpropagating
traveling waves, $p^{\prime }=p_{+}e^{i\left( \omega t-kz\right)
}+p_{-}e^{i\left( \omega t+kz\right) }+c.c.$, where $t$ is time, $z$ is the
axial (propagation) coordinate, $\omega =2\pi f$, and $k=\omega /c_{0}$,
whose complex amplitudes $p_{\pm }$ are related through their reflections at
the boundaries, and the temperature field is decomposed into a homogeneous
component and a grating component, $T^{\prime }=T_{\mathrm{h}}+T_{\mathrm{g}%
}e^{i2kz}+T_{\mathrm{g}}^{\ast }e^{-i2kz}$. All these amplitudes are slowly
varying functions of space and time as the fast (acoustical) variations are
explicitly taken into account through the complex exponentials. Under the
assumption of highly reflecting plates we can adopt a mean field model,
where the slowly varying amplitudes do not depend on the axial coordinate $z$
and $p_{-}=p_{+}\equiv p$, ending up with the following dimensionless
equations \cite{note}%
\begin{align}
\tau _{\mathrm{p}}\partial _{\tau }P& =-P+P_{\mathrm{in}}+i\nabla
^{2}P+i\left( H+G-\Delta \right) P,  \notag \\
\partial _{\tau }H& =-H+D\nabla ^{2}H+2\left\vert P\right\vert ^{2},
\label{dH} \\
\partial _{\tau }G& =-\tau _{\mathrm{g}}^{-1}G+D\nabla ^{2}G+\left\vert
P\right\vert ^{2}.  \notag
\end{align}%
Here $P=\left( \frac{\sigma \omega t_{\mathrm{p}}t_{\mathrm{h}}\alpha _{0}}{%
2\rho _{0}^{2}c_{0}c_{p}}\right) ^{1/2}p$, $H=\omega t_{\mathrm{p}}\sigma T_{%
\mathrm{h}}$,$\;$and $G=\omega t_{\mathrm{p}}\sigma T_{\mathrm{g}}$, are new
normalized variables, $\tau =t/t_{\mathrm{h}}$ is time measured in units of
the relaxation time $t_{\mathrm{h}}$\ of the temperature field homogeneous
component, and $\tau _{\mathrm{p}}=t_{\mathrm{p}}/t_{\mathrm{h}}\;$and $\tau
_{\mathrm{g}}=t_{\mathrm{g}}/t_{\mathrm{h}}$ are the normalized relaxation
times of the intracavity pressure field and the temperature grating
component, respectively. Their original values are given by $t_{\mathrm{p}%
}^{-1}=c_{0}\mathcal{T}/2L+c_{0}\alpha _{0}$, where $\mathcal{T}\ll 1$ is
the transmissivity of the plates and $\alpha _{0}$ is the absorption
coefficient, and $t_{\mathrm{g}}=1/4k^{2}\chi $, where $\chi =\kappa /\rho
_{0}c_{p}$ is the coefficient of thermal diffusivity, $\rho _{0}$ is the
equilibrium density of the medium and $\kappa $ and $c_{p}$ are the thermal
conductivity and the specific heat of the fluid at constant pressure,
respectively. Other parameters are the detuning $\Delta =\left( \omega _{%
\mathrm{c}}-\omega \right) t_{\mathrm{p}}$, with $\omega _{\mathrm{c}}$ the
cavity frequency that lies nearest to the driving frequency $\omega $, and $%
P_{\mathrm{in}}=\frac{c_{0}t_{\mathrm{p}}}{2L}\left( \frac{\sigma \omega t_{%
\mathrm{p}}t_{\mathrm{h}}\alpha _{0}}{2\rho _{0}^{2}c_{0}c_{p}}\right)
^{1/2}p_{\mathrm{in}},$ being $p_{\mathrm{in}}$ the injected pressure plane
wave amplitude, which we take as real without loss of generality. Finally $%
\nabla ^{2}=\frac{\partial ^{2}}{\partial x^{2}}+\frac{\partial ^{2}}{%
\partial y^{2}}$ is the transverse Laplacian operator, where the
dimensionless transverse coordinates $\left( x,y\right) $ are measured in
units of the diffraction length $l_{\mathrm{d}}=c_{0}\sqrt{t_{\mathrm{p}%
}/2\omega }$, and the normalized diffusion coefficient $D=\chi t_{\mathrm{h}%
}/l_{\mathrm{d}}^{2}$.

The model parameters can be estimated for a typical experimental situation 
\cite{Lyakhov93}. We consider a resonator with high quality plates ($%
\mathcal{T}=0.1$), separated by $L=5\unit{cm}$, driven at a frequency $f=2%
\unit{MHz}$, and containing glycerine at 10$\unit{%
{}^{\circ}{\rm C}%
}$. Under these conditions the medium parameters are $c_{0}=2\times 10^{3}%
\unit{m}\unit{s}^{-1}$, $\alpha _{0}=10\unit{m}^{-1}$, $\rho _{0}=1.2\times
10^{3}\unit{kg}\unit{m}^{-3}$, $c_{p}=4\times 10^{3}\unit{J}\unit{kg}^{-1}%
\unit{K}^{-1}$, $\sigma =10^{-2}\unit{K}^{-1}$, and $\kappa =0.5\unit{W}%
\unit{m}^{-1}\unit{K}^{-1}$ ($\chi =10^{-7}\unit{m}^{2}\unit{s}^{-1}$). In
this case $t_{\mathrm{p}}=2\times 10^{-5}\unit{s}$, $t_{\mathrm{g}}=6\times
10^{-2}\unit{s}$, and our length unit $l_{\mathrm{d}}=2\unit{mm}$. For a
resonator with a large Fresnel number, the relaxation of the homogeneous
component of the temperature is mainly due to the heat flux through the
boundaries, and can be estimated from the Newton's cooling law as $t_{%
\mathrm{h}}\sim 10^{1}\unit{s}$. (Remind that this is our time unit.) Then
the diffusion constant $D\sim 10^{0}$, and the normalized decay times $\tau
_{\mathrm{p}}\sim 10^{-6}$, and$\;\tau _{\mathrm{g}}\sim 10^{-2}$ under
usual conditions. We see that the problem is typically very stiff: $0<\tau _{%
\mathrm{p}}\ll \tau _{\mathrm{g}}\ll 1$. In the following the results will
be given to the lowest nontrivial order in these smallest decay times in
order to not overburdening the expressions.

The spatially uniform steady state can be obtained by neglecting the
derivatives in Eqs. (\ref{dH}). Introducing the notation $W=\left\vert
P\right\vert ^{2}$, $W_{\mathrm{in}}=\left\vert P_{\mathrm{in}}\right\vert
^{2}$, one has $G=\tau _{\mathrm{g}}W$,$\;H=2W$, and%
\begin{equation}
W_{\mathrm{in}}=W+\left( \Delta -2W\right) ^{2}W.  \label{YX}
\end{equation}%
The characteristic curve $W$ vs. $W_{\mathrm{in}}$ can display an S-shape,
typical of the optical bistability of coherently driven optical Kerr
cavities, as we show next. The two turning points of the characteristic ($W_{%
\mathrm{in,}\pm },W_{\pm }$), verifying $\mathrm{d}W_{\mathrm{in}}/\mathrm{d}%
W=0$, and its inflection point ($W_{\mathrm{in,I}},W_{\mathrm{I}}$), defined
by $\mathrm{d}^{2}W_{\mathrm{in}}/\mathrm{d}W^{2}=0$, are given by $W_{\pm }=%
\frac{2\Delta \pm \sqrt{\Delta ^{2}-3}}{6}$ and $W_{\mathrm{I}}=\frac{\Delta 
}{3}$, and the corresponding values of the input pressure follow from Eq. (%
\ref{YX}). Note that bistability requires $\Delta >\Delta _{0}=\sqrt{3}$,
the nascent bistability (NB) occurring at $\Delta =\Delta _{0}$, in which
case $W_{\pm }=W_{\mathrm{I}}=W_{0}=\frac{1}{\sqrt{3}}$. It is very easy to
show that this NB inflection point occurs at $P_{\mathrm{in},0}=\frac{2}{%
3^{3/4}}$ and that the values of the model variables become $P_{0}=\frac{%
\sqrt{3}-i}{2\cdot 3^{1/4}}$, $H_{0}=2W_{0}$, and $G_{0}=\tau _{\mathrm{g}%
}W_{0}$. Figure 1 shows the behaviour of the homogeneous steady state
illustrating the effect of detuning on the character of the solutions \cite%
{bistability}.

\begin{figure}[h]
\centering\includegraphics[width=0.3\textwidth]{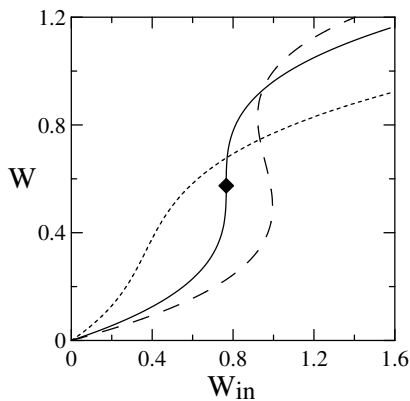}
\caption{{}Characteristic curve output intensity vs input intensity as
following from Eq. (\protect\ref{YX}), for three values of the detuning $%
\Delta $. Monostable regime (dotted line) for $\Delta =1$, nascent
bistability (full line) for $\Delta =\protect\sqrt{3}$, and bistable regime
(dashed line) for $\Delta =2$ are shown$.$ The symbol marks the position of
the inflection point ($X_{0},Y_{0}$) at NB.}
\end{figure}

Inspired by previous works on passive optical cavities \cite{1OBpatterns,2OB}
we shall concentrate our study on conditions close to nascent bistability ($%
\Delta \simeq \Delta _{0}$) and to the vicinity of the inflection point, $%
\left( P,H,G,P_{\mathrm{in}}\right) \simeq \left( P_{0},H_{0},G_{0},P_{%
\mathrm{in,}0}\right) $, where interesting pattern formation properties can
be envisaged. Let us just comment before considering in detail the
spatiotemporal dynamics of the thermoacoustic resonator that, by focusing on
the above conditions we can describe in a consistent way the S-shape of the
characteristic curve by using standard multiple scale methods \cite{Nayfeh}.

We consider next the stability of the spatially uniform steady state. We
consider perturbations of the form $\exp \left( i\mathbf{k}\cdot \mathbf{r}%
+\lambda t\right) $ and linearize the model Eqs. (\ref{dH}) with respect to
them. Further considering the relevant case $0<\tau _{\mathrm{p}}\ll \tau _{%
\mathrm{g}}\ll 1$, see typical values above, one obtains that, to the
leading order, the Lyapunov exponent $\lambda $ can take on the following
values: $\lambda \in \left\{ -\tau _{\mathrm{p}}^{-1},-\tau _{\mathrm{p}%
}^{-1},-\tau _{\mathrm{g}}^{-1},\lambda _{\mathrm{+}}\right\} $, where%
\begin{equation}
\lambda _{\mathrm{+}}=-1-Dk^{2}+4W\frac{\Delta +k^{2}-2W}{1+\left( \Delta
+k^{2}-2W\right) ^{2}},  \label{lambda}
\end{equation}%
$k=\left\vert \mathbf{k}\right\vert $, and we remind that $W=\left\vert
P\right\vert ^{2}$, which fixes the injection pressure through Eq. (\ref{YX}%
). The obtention of an order parameter equation through a multiple scales
analysis requires that a small eigenvalue exists at small wavenumber $k$. In
our case $\lambda _{\mathrm{+}}$ is that eigenvalue. For small $k$ it reads $%
\lambda _{\mathrm{+}}=a+bk^{2}+\mathcal{O}\left( k^{4}\right) $, $a=4W\frac{%
\Delta -2W}{1+\left( \Delta -2W\right) ^{2}}-1$, $b=\frac{8W}{\left[
1+\left( \Delta -2W\right) ^{2}\right] ^{2}}-\frac{4W}{1+\left( \Delta
-2W\right) ^{2}}-D$. As we will be concerned with the neighbourhood of the
inflection point under NB conditions, let us particularize our analysis to
that point. In such case $W=W_{0}$, $\Delta =\Delta _{0}$ and we obtain $a=0$
and $b=D_{0}-D$, $D_{0}=\frac{\sqrt{3}}{2}$. Hence, $\lambda _{\mathrm{+}}$
is actually small for small $k$ in the region of interest. Finally, as we
wish to capture pattern formation asymptotically, we must impose that $%
\lambda _{+}$ can change its sign for small $k$. This imposes that the
diffusion coefficient $d=D-D_{0}\sim \mathcal{O}(\varepsilon )$, being $%
\varepsilon $ a small parameter. Additionally, a careful analysis of the
homogeneous steady state reveals that, in order to describe consistently the
vicinity of the NB point, one must consider the scalings $\delta =\Delta
-\Delta _{0}\sim \mathcal{O}(\varepsilon ^{2})$ and $\mu =\frac{P_{\mathrm{in%
}}-P_{\mathrm{in},\mathrm{I}}}{2\cdot 3^{1/4}}\sim \mathcal{O}(\varepsilon
^{3})$, with $P_{\mathrm{in},\mathrm{I}}=\frac{2}{3^{3/4}}\left( 1+\frac{%
\delta }{2}\right) $. Finally the multiple scale analysis can be performed
assuming $p=P-P_{0}\sim \mathcal{O}(\varepsilon )$, $h=H-H_{0}\sim \mathcal{O%
}(\varepsilon )$ and $g=G-G_{0}\sim \mathcal{O}(\varepsilon ),$ together
with the slow space and time scales $\tau \sim \mathcal{O}(\varepsilon
^{-2}) $ and $\nabla ^{2}\sim \mathcal{O}(\varepsilon )$. Following standard
procedures \cite{Nayfeh}, a closed equation for $h$ is obtained as a
solvability condition at the third order in $\varepsilon $, which reads%
\begin{eqnarray}
\partial _{\tau }h &=&\mu +\tfrac{\sqrt{3}\delta }{2}h-\tfrac{3}{4}%
h^{3}+d\nabla ^{2}h-\tfrac{3}{2}\nabla ^{4}h  \notag \\
&&-\tfrac{3}{4}h\nabla ^{2}h-\tfrac{3}{2}\nabla ^{2}h^{2}.  \label{OPE}
\end{eqnarray}%
The other fields are related to $h$ as $g=\frac{\tau _{\mathrm{g}}}{2}h$,
and $p=\frac{3^{1/4}\left( \sqrt{3}+i\right) }{4}h$ to the leading order.
Equation (\ref{OPE}) is a main result of this Letter as it describes the
spatiotemporal dynamics of the thermoacoustic resonator in a very simple
way. Moreover this equation can be considered as "universal" for our system
under typical operating conditions as the numerical values of the
coefficients appearing in it are, in the considered limit $\tau _{\mathrm{g}%
}\rightarrow 0$, very weakly dependent on $\tau _{\mathrm{g}}$ (details will
be given elsewhere). Equation (\ref{OPE}) is a modified Swift-Hohenberg
equation, and has been previously introduced in the optical context for
semiconductor lasers \cite{Kozyreff} and liquid crystal light valve (LCLV)
systems \cite{Durniak05,Clerc05}. In this context the analysis of Eq. (\ref%
{OPE}) has demonstrated different pattern formation scenarios, including
front propagation, pattern competition and localized structures.
Experimental evidence has been also reported in \cite{Clerc05}. Despite the
fundamental differences between the thermoacoustic resonator considered here
and the referred optical systems, the derivation of a common order parameter
equation allows to extend the optical predictions to the acoustic case.

The homogeneous steady solution $h_{0}$ of Eq. (\ref{OPE}) is implicitly
given by $\mu =\frac{3}{4}h_{0}^{3}-\tfrac{\sqrt{3}}{2}\delta h_{0}$, being
multivalued when $\delta >0$ $(\Delta >\Delta _{0}),$ in agreement with the
analysis of the microscopic model. In the following we focus on the bistable
region, extending between the turning points $h_{0,\pm }=\pm \frac{\sqrt{%
2\delta }}{3^{3/4}}$. A linear stability analysis shows that, for $d>0$,
only the upper branch can be modulationally unstable \cite{Durniak05},
giving rise to a pattern with wavenumber $k_{\mathrm{c}}^{2}=\frac{15h_{0}-4d%
}{12}$. These results are confirmed by the numerical integration of Eq. (\ref%
{OPE}), performed in a one-dimensional case ($\nabla =\partial _{x}^{2}$)
which describes, e.g. a physical situation in which the resonator has a slab
waveguide configuration confining the sound in the $y$ direction. Figure
2(a) shows the spatiotemporal evolution (from bottom to top) of an initially
homogeneous distribution, where the developement of a modulational
instability is observed, in agreement with the previous analysis. 
\begin{figure}[h]
\centering\includegraphics[width=0.45\textwidth]{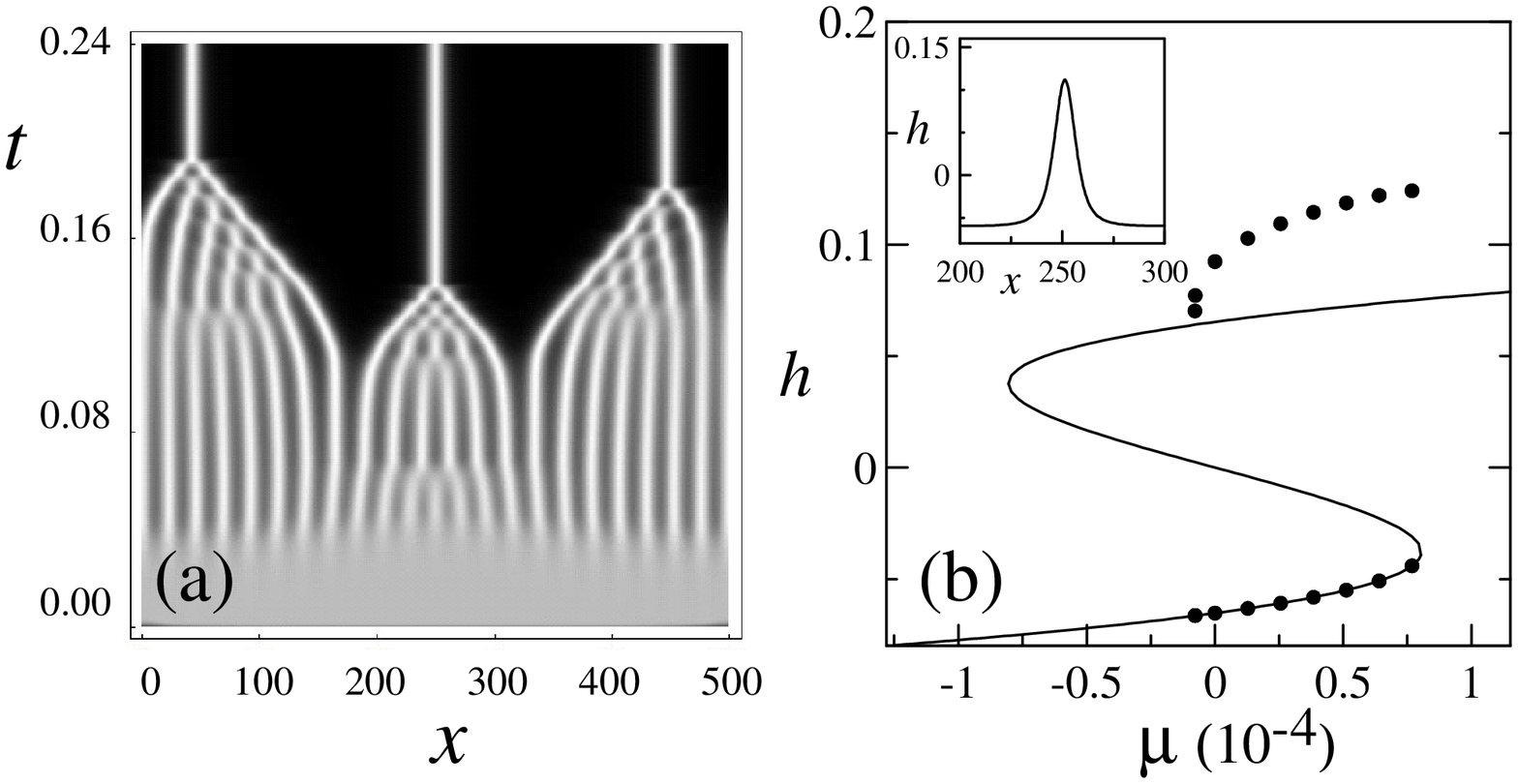}
\caption{(a) Numerical simulation of Eq. (\protect\ref{OPE}) for $\protect%
\delta =3~10^{-3}.$ (a) The modulational instability and the further
collapse resulting in an array of CSs is shown for $\protect\mu =3~10^{-5}$.
Time runs from bottom to top. (b) Maximum amplitude and background values of
the CSs (dots), together with the homogeneous steady solution (full line)
depending on the control parameter $\protect\mu $. The inset shows the
profile of a CS for parameters in Fig. 2(a). All variables are dimensionless.
}
\end{figure}

In most of the multivalued domain, the extended patterns emerging at the
modulational instability are unstable, being a transient state. As shown in
Fig. 2(a), neighbour maxima collide and merge, the long-term evolution
resulting in a number of localized structures or CSs. In Fig. 2(b) the
background and peak values of the CSs (see a profile in the inset) depending
on the control parameter $\mu $ are shown with dots. The background
amplitude corresponds to the stable uniform state, while the peak value to
that of the underlying extended pattern which, as stated in the
introduction, is a signature of CSs. Note that CSs are stable and robust
structures whenever a pattern and a uniform state coexist.

Figure 2 reveals some important facts. Firstly, localized structures can
form spontaneously from the homogeneous steady state, differently from other
systems which require a hard local excitation of the medium. Such
circumstance is very relevant in the particular case considered here, where
the nature of the driving source (a plane, vibrating rigid surface) makes
difficult to implement experimentally a localized excitation. Certainly, the
proposal of techniques to address CS in the acoustic resonator will be most
relevant.

We have also obtained an analytical, steady CS of Eq. (\ref{OPE}),%
\begin{equation}
h(x)=h_{0}+h_{\mathrm{s}}\func{sech}^{2}(x/x_{\mathrm{s}}),  \label{soliton}
\end{equation}%
where the background $h_{0}$ corresponds to the stable homogeneous solution
of lower amplitude. The expressions of the peak amplitude $h_{\mathrm{s}}$
and width $x_{\mathrm{s}}$ in (\ref{soliton}) are quite involved but can be
readily obtained substituting Eq. (\ref{soliton}) into Eq.(\ref{OPE}). The
solution given by Eq. (\ref{soliton}) corresponds to an ultrasonic CS but
also applies to optical CSs in those systems described by Eq. (\ref{OPE}).
It is a singular solution, existing for a special value of the injected
amplitude $\mu $ determined by the rest of parameters $\left( d,\delta
\right) $, which is seen to be a particular case of the (non-analytic) CSs
family found in the numerical study and illustrated in Fig. 2(b).

Concluding, we have studied the spatiotemporal dynamics of ultrasound in a
resonator containing a viscous medium --a thermoacoustic resonator. The
microscopic equations have been reduced, close to the nascent bistability
point, to a single order parameter equation previously obtained in other
contexts. The analysis of the reduced model demonstrates the existence of
bistability and modulational instabilities. As a consequence, the system is
shown to support cavity solitons, corresponding to states where ultrasound
is highly localized in the transverse plane of the resonator. Analytical
solutions of such ultrasonic CSs have been obtained. The results reported
here will also find application to those systems described by the
generalized Swift-Hohenberg equation (\ref{OPE}), which is a generic model
for pattern formation.

The work was financially supported by the Spanish Ministerio de Educaci\'{o}%
n y Ciencia and the European Union FEDER (Projects FIS2005-07931-C03-01 and
-02, and Programa Juan de la Cierva).


\begin{thebibliography}{99}
\bibitem{Cross93} M. C. Cross and P. C. Hohenberg, Rev. Mod. Phys. \textbf{65%
}, 851 (1993).

\bibitem{Newell93} A. C. Newell, T. Passot and J. Lega, Ann. Rev. Fluid
Mech. \textbf{25}, 399 (1993).

\bibitem{NM92} A. C. Newell and J. V. Moloney, \textit{Nonlinear optics}
(Addison-Wesley, 1992).

\bibitem{Mandel97} P. Mandel, \textit{Theoretical problems in cavity
nonlinear optics} (Cambridge University Press, 1997).

\bibitem{Springer03} K. Staliunas and V. J. S\'{a}nchez-Morcillo, \textit{%
Transverse patterns in nonlinear optical resonators} (Springer, 2003).

\bibitem{Barland02} S. Barland \textit{et al.}, Nature \textbf{419}, 699
(2002).

\bibitem{Bunkin86} F. V. Bunkin, Yu. A. Kravtsov, and G. A. Lyakhov, Sov.
Phys. Usp. \textbf{29}, 607\ (1986).

\bibitem{Bunkin94} F. V. Bunkin, G. A. Lyakhov, and K. F. Shipilov, Phys.
Usp. \textbf{38}, 1099 (1995).

\bibitem{Naugolnykh} K. Naugolnykh and L. A. Ostrovsky, \textit{Nonlinear
wave processes in acoustics} (Cambridge University Press, 1998).

\bibitem{Lyakhov93} G. A. Lyakhov \textit{et al.}, Acoust. Phys. \textbf{39}%
, 158 (1993).

\bibitem{Andreev86} V. G. Andreev \textit{et al.}, Sov. Phys. Acoust.\textbf{%
\ 32}, 404 (1986).

\bibitem{note} Details of the derivation will be given elsewhere.

\bibitem{bistability} Bistability actually requires simultaneous stability
of two solutions, what can occur in our model as we show below. Nevertheless
here the important feature is the multivaluedness of the characteristic
around the inflection point.

\bibitem{1OBpatterns} M. Tlidi, M. Georgiou, and P. Mandel, Phys. Rev. A 
\textbf{48}, 4605 (1993); M. Tlidi, P. Mandel, and R. Lefever, Phys. Rev.
Lett. \textbf{73}, 640 (1994).

\bibitem{2OB} V. J. S\'{a}nchez-Morcillo and G. J. de Valc\'{a}rcel, Quantum
Semiclass. Opt. \textbf{8}, 919 (1996).

\bibitem{Nayfeh} A. H. Nayfeh, \textit{Perturbation methods} (John Wiley \&
Sons, New York, 2000)

\bibitem{Kozyreff} G. Kozyreff, S. J. Chapman, and M. Tlidi, Phys. Rev. E%
\textbf{\ 68}, 015201(R) (2003); G. Kozyreff and M. Tlidi, Phys. Rev. E 
\textbf{69}, 066202 (2004).

\bibitem{Durniak05} C. Durniak \textit{et al.}, Phys. Rev. E\textbf{\ 72},
026607 (2005).

\bibitem{Clerc05} M.\ G. Clerc, A. Petrossian, and S. Residori, Phys. Rev.
E, \textbf{71, }015205 (R) (2005).
\end{thebibliography}
\end{document}